\documentclass[aps,prd,preprintnumbers,twocolumn,superscriptaddress,nofootinbib]{revtex4-2}
\usepackage{lmodern}
\usepackage{hyperref}
\hypersetup{colorlinks=true,linkcolor=purple,anchorcolor=blue,citecolor=blue, filecolor=blue,urlcolor=blue,bookmarksnumbered=true,
pdfview=FitB
}
\usepackage{mathrsfs}
\usepackage{rsfso}
\usepackage{color}
\usepackage{xcolor}
\usepackage{ulem}
\usepackage{csquotes}
\colorlet{purple1}{blue!70!red}
\colorlet{darkred}{red!50!black}

\usepackage{graphicx}
\usepackage{psfrag}
\usepackage{color}
\usepackage{amssymb}
\usepackage{amsmath}
\usepackage{epstopdf}
\usepackage{natbib}
\usepackage{amssymb}
\usepackage{amsmath,amssymb}
\usepackage{mathtools}
\usepackage{float}
\usepackage{color}
\usepackage{leftidx}
\usepackage{booktabs}
\usepackage{latexsym}
\usepackage{revsymb}
\usepackage{multirow}
\usepackage{hypcap}
\usepackage{array}
\usepackage[english]{babel}
\usepackage{amsmath}
\usepackage{cleveref}
\def\orcid#1{\kern .08em\href{https://orcid.org/#1}{\includegraphics[keepaspectratio,width=0.7em]{ORCID_iD.png}}}
\newcommand{\be}{\begin{eqnarray}}
	\newcommand{\ee}{\end{eqnarray}}
	
\def\orcid#1{\kern .08em\href{https://orcid.org/#1}{\includegraphics[keepaspectratio,width=0.7em]{ORCID_iD.png}}}

\usepackage{graphicx}
\usepackage{bm}

\usepackage{subcaption}
\usepackage{mathrsfs}
\usepackage{float}
\usepackage{url}
\usepackage{ulem}
\graphicspath{{images}}

\allowdisplaybreaks

\begin{document}


\title{Dynamical gluon effects in twist-3 generalized parton distributions of the proton}

\author{Ziqi Zhang}
\email{zhangziqi@impcas.ac.cn}
\affiliation{Institute of Modern Physics, Chinese Academy of Sciences, Lanzhou 730000, China}
\affiliation{School of Nuclear Science and Technology, University of Chinese Academy of Sciences, Beijing 100049, China}
\affiliation{CAS Key Laboratory of High Precision Nuclear Spectroscopy, Institute of Modern Physics, Chinese Academy of Sciences, Lanzhou 730000, China}

\author{Chandan Mondal}
\email{mondal@impcas.ac.cn}
\affiliation{Institute of Modern Physics, Chinese Academy of Sciences, Lanzhou 730000, China}
\affiliation{School of Nuclear Science and Technology, University of Chinese Academy of Sciences, Beijing 100049, China}
\affiliation{CAS Key Laboratory of High Precision Nuclear Spectroscopy, Institute of Modern Physics, Chinese Academy of Sciences, Lanzhou 730000, China}

\author{Siqi Xu}
\email{xsq234@impcas.ac.cn}
\affiliation{Institute of Modern Physics, Chinese Academy of Sciences, Lanzhou 730000, China}
\affiliation{School of Nuclear Science and Technology, University of Chinese Academy of Sciences, Beijing 100049, China}
\affiliation{CAS Key Laboratory of High Precision Nuclear Spectroscopy, Institute of Modern Physics, Chinese Academy of Sciences, Lanzhou 730000, China}

\author{Xingbo Zhao}
\email{xbzhao@impcas.ac.cn}
\affiliation{Institute of Modern Physics, Chinese Academy of Sciences, Lanzhou 730000, China}
\affiliation{School of Nuclear Science and Technology, University of Chinese Academy of Sciences, Beijing 100049, China}
\affiliation{CAS Key Laboratory of High Precision Nuclear Spectroscopy, Institute of Modern Physics, Chinese Academy of Sciences, Lanzhou 730000, China}

\author{James P. Vary}
\email{jvary@iastate.edu}
\affiliation{Department of Physics and Astronomy, Iowa State University, Ames, IA 50011, U.S.A.}

\collaboration{BLFQ Collaboration}

\date{\today}

\begin{abstract}

Within the Basis Light-Front Quantization framework, we systematically investigate the subleading-twist (twist-3) generalized parton distributions (GPDs) of the proton’s valence quarks beyond the Wandzura--Wilczek (WW) approximation. The twist-3 GPDs are not independent; through the equations of motion they decompose into a non-genuine contribution and a genuine twist-3 term. The latter encodes quark--quark--gluon correlations and involves interference between the light-front Fock sectors $|qqq\rangle$ and $|qqqg\rangle$, which are typically neglected in the WW approximation. Using light-front wave functions obtained from diagonalizing the proton light-front Hamiltonian for its $|qqq\rangle$ and $|qqqg\rangle$ Fock components, we compute these GPDs via their overlap representations. To further explore their physical implications, we also evaluate several twist-3--related quantities, including the quark orbital angular momentum, the total quark spin contribution, and the quark spin--orbit correlation. Our results provide new nonperturbative input on higher-twist dynamics particularly multi-parton interference effects relevant for future measurements at the EicC and the EIC.

\end{abstract}

\maketitle


\section{\label{sec:level1}Introduction}
One of the most intriguing problems in modern physics is the exploration of the proton’s internal structure and its spin puzzle, both experimentally and theoretically (see Refs.~\cite{Diehl:2023nmm,Lorce:2025aqp} and references therein). A powerful tool for probing the proton structure is deeply virtual Compton scattering (DVCS), in which a quark absorbs a virtual photon emitted by an electron, promptly emits a real photon, and then returns to the proton. This process has been measured in numerous experiments, including JLab CLAS~\cite{girod2008measurement,pisano2015single,jo2015cross,hattawy2019exploring}, JLab Hall~A~\cite{camacho2006scaling,mazouz2007deeply}, HERA H1~\cite{adloff2001measurement,aaron2008measurement,aaron2009deeply}, HERA ZEUS~\cite{adloff2001measurement,zeus2009measurement}, and HERA HERMES~\cite{airapetian2012beam,airapetian2012beam2}. Further measurements are anticipated at the future Electron–Ion Collider (EIC)~\cite{khalek2022science} and the Electron–Ion Collider in China (EicC)~\cite{anderle2021electron}.

A variety of distribution functions describe the proton’s internal structure at the partonic level. For hard exclusive processes (e.g., DVCS), generalized parton distributions (GPDs) are employed, providing access to the proton’s three-dimensional structure. GPDs have attracted significant interest because they are related to conventional observables while encoding much richer information. In the forward limit ($\Delta \rightarrow 0$), one recovers the ordinary parton distribution functions (PDFs), which describe the longitudinal momentum distributions of quarks and gluons in the proton. Conversely, integrating GPDs over the longitudinal momentum fraction yields the electromagnetic form factors (FFs), which characterize the proton’s charge and magnetization distributions.

Several theoretical studies of proton GPDs have been carried out using a wide range of QCD-inspired models, including the bag model~\cite{Ji:1997gm,Anikin:2001zv}, soliton models~\cite{Goeke:2001tz,Petrov:1998kf,Penttinen:1999th}, constituent quark models (CQMs)~\cite{Scopetta:2003et,Scopetta:2002xq,Boffi:2002yy,Boffi:2003yj,Scopetta:2004wt}, light-front (LF) quark–diquark models~\cite{Mondal:2017wbf,Maji:2017ill,Chakrabarti:2015ama,Maji:2015vsa,Mondal:2015uha}, meson-cloud models~\cite{Pasquini:2006dv,Pasquini:2006ib}, and approaches based on AdS/QCD~\cite{Vega:2010ns,Chakrabarti:2013gra,Rinaldi:2017roc,Traini:2016jko,deTeramond:2018ecg,Gurjar:2022jkx}. Additional progress has been made within LF quantization frameworks~\cite{Tiburzi:2001ta,Tiburzi:2001je,Mukherjee:2002xi,Lin:2023ezw,Lin:2024ijo,zhang2024twist}. At the same time, Euclidean lattice QCD has emerged as a promising first-principles framework for computing GPDs~\cite{Ji:2013dva,Ji:2020ect,Lin:2021brq,Lin:2020rxa,Bhattacharya:2022aob,Alexandrou:2021bbo,Alexandrou:2022dtc,Guo:2022upw,Alexandrou:2020zbe,Gockeler:2005cj,QCDSF:2006tkx,Alexandrou:2019ali}.

However, most of the studies mentioned above focus on the leading-twist (twist-2) level. Twist-3 power counting of physical observables are often assumed to be small because they are suppressed by a factor of \(1/Q\), where \(Q\) denotes the four-momentum transfer of the process. In reality, twist-3 effects are non-negligible and play an important role in revealing the proton’s internal structure~\cite{zhang2024twist,tan2024chiral,jain2024deciphering,Vladimirov:2025qrh}. First, twist-3 distributions are directly connected to quark orbital angular momentum (OAM) and quark spin--orbit correlations~\cite{ji1997gauge,hatta2012twist,JAFFE1990509}. Second, they are related to the average transverse color Lorentz force experienced by quarks~\cite{PhysRevD.88.114502,aslan2019transverse}. Third, twist-3 GPDs at nonzero skewness enter the twist-3 DVCS amplitude through the corresponding Compton form factors (CFFs)~\cite{guo2022twist}.



In this work, we employ the theoretical framework of Basis Light-Front Quantization (BLFQ)~\cite{Vary:2009gt} and use light-front wave functions (LFWFs), obtained by diagonalizing the LFQCD Hamiltonian, to calculate all twist-3 GPDs in the zero-skewness limit. In contrast to previous studies~\cite{zhang2024twist}, our approach extends the Fock-space truncation to include the first two sectors, $|qqq\rangle$ and $|qqqg\rangle$, thereby incorporating contributions from quark--gluon--quark correlations. This goes beyond the Wandzura--Wilczek (WW) approximation~\cite{kivel2001wandzura,radyushkin2000dvcs,kivel2001dvcs,radyushkin2001kinematical}. We further explore the connections between twist-3 GPDs, quark OAM, and quark spin--orbit correlations.

The structure of this work is as follows: Sec. \ref{sec2} provides a brief introduction to the BLFQ framework; Sec. \ref{sec3} outlines GPDs and presents the twist decomposition of GPDs; Sec. \ref{sec4} reports numerical results for twist-3 zero-skewness GPDs within the first two Fock sectors, along with their properties in the forward limit; Sec. \ref{sec5} summarizes the work and discusses future prospects.

\section{\label{sec:level1}Basis Light-front Quantization} \label{sec2}
As a non-perturbative approach, BLFQ is grounded in the Hamiltonian formalism and leverages the advantages of LF dynamics. Its core idea is to simultaneously obtain the mass spectrum and bound state wavefunctions by solving the eigenvalue problem 
\begin{equation}
    P^- P^+ |\psi \rangle = M^2 |\psi \rangle ,
\end{equation}
where $P^+$ and $P^-$ denote the longitudinal momentum and the LF Hamiltonian of the system, respectively, with the `$\pm$' components being defined as $P^{\pm} = P^0 \pm P^3$. Here, $M^2$ is the eigenvalue corresponding to the LF energy of the system, and the eigenvector $|\psi \rangle$, representing the proton state, is expanded as
\begin{equation}
    |\psi_{\mathrm{proton}} \rangle = \Phi^{qqq} |qqq \rangle + \Phi^{qqqg} |qqqg \rangle + \cdots ,
\end{equation}
where $\Phi$ are the corresponding LFWFs, and the `$\cdots$' represents all other possible partonic combinations that can be found inside the proton.

At the initial scale, where the proton state is described by the $|qqq\rangle$ and $|qqqg\rangle$ Fock components, we employ the LF Hamiltonian
$P^- = P^-_{\rm QCD} + P^-_{\rm C}$,
where $P^-_{\rm QCD}$ incorporates the relevant QCD interactions and $P^-_{\rm C}$ models confinement~\cite{Xu:2023nqv}.
In the LF gauge $A^+ = 0$, the LF QCD Hamiltonian with one dynamical gluon is given by \cite{Xu:2023nqv}
\begin{align} \label{pqcd}
    P^-_{\mathrm{QCD}} =& \int d^2x^\perp dx^- \bigg\{\frac{1}{2} \bar{\psi} \gamma^+ \frac{m_0^2 + (i\partial^\perp)^2}{i\partial^+} \psi \notag \\
    & +\frac{1}{2} A_a^i \left[ m_g^2 + (i\partial^\perp)^2 \right] A_a^i 
    + g_s \bar{\psi} \gamma_\mu T^a A_a^\mu \psi \notag \\
    & +\frac{1}{2} g_s^2 \bar{\psi} \gamma^+ T^a \psi \frac{1}{(i\partial^+)^2} \bar{\psi} \gamma^+ T^a \psi \bigg\},
\end{align}
where the first and second terms \footnote{The sign of the second term of Eq. (2) in Ref. \cite{Xu:2023nqv} should be $+$.} account for the kinetic energies of the quark (with bare mass \( m_0 \)) and gluon (with bare mass \( m_g \)), respectively, while the last two terms represent the vertex and instantaneous interactions, characterized by the strong coupling \( g_s \). Following the spirit of Ref. \cite{Burkardt:1998dd}, we introduce an independent vertex mass $m_f$ to parameterize nonperturbative effects related to chiral symmetry breaking within light-front dynamics. A detailed description and physical interpretation of these terms can be found in Refs.~\cite{Xu:2023nqv,Zhu:2024awq,nair2025proton}.

Confinement in the leading Fock sector is implemented following Ref.~\cite{Li:2015zda} as
\begin{equation}
    P^-_{\rm C} P^+ = \frac{\kappa^4}{2} \sum_{i\neq j} \left[ \vec{r}_{ij\perp}^{\,2}
    - \frac{\partial_{x_i}\!\left(x_i x_j \partial_{x_j}\right)}{(m_i + m_j)^2} \right],
\end{equation}
where $\vec{r}_{ij\perp} = \sqrt{x_i x_j}(\vec{r}_{i\perp}-\vec{r}_{j\perp})$ denotes the transverse separation between quarks $i$ and $j$, while $\kappa$ is the confinement strength.  
In the $|qqqg\rangle$ sector, we omit an explicit confinement term, assuming that the restricted transverse basis together with the inclusion of a massive gluon effectively captures the essential confinement behavior.

The many-particle basis includes the transverse center-of-mass (CM) motion that is entangled with the intrinsic motion. Consequently, it is essential to introduce a constraint term \cite{Vary:2009gt} into the effective Hamiltonian
\begin{equation}
    H^\prime = \lambda_L (H_{\mathbf{cm}} - 2b^2 I),
\end{equation}
where $\lambda_L$ denotes the Lagrange multiplier, $2b^2$ represents the zero-point energy, and $I$ is the identity operator. By choosing a sufficiently large $\lambda_L$, one can shift the excited states of the CM motion to higher energies and ensure that all low-lying states correspond to the ground state of the CM motion.

We solve the LF Hamiltonian eigenvalue problem within the BLFQ framework~\cite{Vary:2009gt}, expanding the proton in a basis of longitudinal plane waves (in a box of length $2L$ with antiperiodic/periodic boundary conditions for quarks/gluons), transverse two-dimensional harmonic oscillator (2D-HO) functions $\Phi_{nm}(\vec{p}_\perp;b)$~\cite{Zhao:2014xaa}, and light-cone helicity spinors. Single-particle states are labeled by $\bar{\alpha}=\{k,n,m,\lambda\}$, with $k$ being the longitudinal
quantum number (half-integer for quarks, integer for gluons, excluding zero mode); $(n,m)$ are the 2D-HO quantum numbers, and $\lambda$ the helicity; states in sectors with multiple color-singlets such as $|qqqg\rangle$ require an additional label to distinguish their color singlet configuration.

The basis is truncated by $N_{\rm max}$ and ${K}=\sum_i k_i$ with the sum running over all $i$-partons from 1 to $N$, $N_{\rm max}$ imposes $\sum_i (2n_i+|m_i|+1)\le N_{\rm max}$, while ${K}$ fixes the longitudinal resolution, with $x_i=k_i/{K}$. These parameters set the IR/UV scales~\cite{Zhao:2014xaa}. Diagonalization yields the proton LFWFs with helicity $\Lambda$,
\begin{equation}
    \Psi^{N,\Lambda}_{\{x_i,\vec{p}_{\perp i},\lambda_i\}}
    = \sum_{\{n_i,m_i\}} \psi^{N}(\{\alpha_i\})
      \prod_{i=1}^{N} \Phi_{n_i m_i}(\vec{p}_{i\perp}, b),
\end{equation}
where $\psi^{N=3}$ and $\psi^{N=4}$ are the eigenvector components for the $|uud\rangle$ and $|uudg\rangle$ sectors.

The Hamiltonian parameters, summarized in Table \ref{paratabel} with $\{N_{\rm{max}},K\} = \{9, 16.5\}$, are set to reproduce the proton mass and its electromagnetic properties~\cite{Xu:2023nqv}. The parameter \(b_{\rm inst}\) sets the regularization scale for the transverse momentum of the instantaneous gluon, with this regulating function reconciling the UV cutoffs between instantaneous and dynamical gluon interactions~\cite{Wu:2026gul}. At the model scale, the proton contains approximately $44\%$ probability in the $|qqq\rangle$ sector and $56\%$ in the $|qqqg\rangle$ sector. The resulting LFWFs, appropriate for a low-resolution scale of $\mu_0^2 \sim 0.24 \pm 0.01~\text{GeV}^2$~\cite{Xu:2023nqv}, have been successfully used to describe a broad range of proton observables, including electromagnetic form factors, radii, PDFs, GPDs, TMDs, and spin and orbital angular momentum~\cite{Xu:2023nqv,Yu:2024mxo,Zhu:2024awq,Zhang:2025nll,Lin:2024ijo,Lin:2023ezw,nair2025proton}.


\begin{table}[h!]
\caption{\label{paratabel}
Summary of the model parameters~\cite{Xu:2023nqv}. All parameters have units of GeV except for $g_s$.}
\begin{ruledtabular}
\begin{tabular}{cccccccc}
$m_u$ & $m_d$ & $m_g$ & $\kappa$ & $m_f$ & $g_s$ & $b$ & $b_{\textbf{inst}}$ \\
\colrule
0.31 & 0.25 & 0.50 & 0.54 & 1.80 & 2.40 & 0.70 & 3.00 \\
\end{tabular}
\end{ruledtabular}
\end{table}

\section{\label{sec:level1}Generalized Parton Distribution} \label{sec3}
In this section, we briefly introduce the definition of twist-3 GPDs. One of the most common definitions is formulated via a bilocal quark-quark correlator, which depends on three variables: \(x\), \(\xi\), and \(-t\). Here, \(x\) denotes the longitudinal momentum fraction, \(\xi\) represents the skewness, and \(-t\) stands for the squared momentum transfer. The correlator takes the form
\begin{align}
    &F^{[\Gamma]}_{\Lambda'\Lambda} (x,\xi,-t) = \frac{1}{2} \int \frac{dy^{-}}{4\pi} e^{ixP\cdot y} \bigg\langle P',\Lambda' \bigg| \bar{\Psi}\left(-\frac{y}{2}\right)  \notag \\
    &\quad\quad\quad\quad\times \mathscr{W}\left(-\frac{y}{2},\frac{y}{2}\right)\Gamma \Psi\left(\frac{y}{2}\right) \bigg| P,\Lambda \bigg\rangle \bigg|_{y^{+}=0,\vec{y}^{\perp}=0} , \label{gpddef}
\end{align}
where \(| P,\Lambda \rangle\) describes the momentum and helicity of the initial proton state, while the primed bra vector (\(\langle P',\Lambda' |\)) corresponds to the final proton state. The Wilson line is defined as \(\mathscr{W}(y_2,y_1) \equiv P \mathrm{exp}(ig_s\int_{y_1}^{y_2} dz^\mu A_\mu(z))\); under the light-cone gauge (\(A^+=0\)), this Wilson line reduces to unity. \(\Gamma\) represents one element of the complete set of Dirac matrices, and \(\Psi(x)\) represents the quark field. As we will see, not all components of \(\Psi(x)\) are independent dynamical degrees of freedom.

By applying a projection operator \(\mathcal{P}^\pm\), the quark field can be decomposed into dynamical and non-dynamical components~\cite{kogut1970quantum}, i.e., \(\Psi = \Psi^+ + \Psi^-\). Here, \(\Psi^\pm = \mathcal{P}^\pm \Psi\) with \(\mathcal{P}^\pm = \gamma^\mp \gamma^\pm / 4\). The dynamical component \(\Psi^+\) is expanded in the form of a free fermion field, while the non-dynamical component \(\Psi^-\) is constrained by the Dirac equation,
\begin{equation}
    \Psi^- = \frac{\gamma^+}{2i\partial^+}\left(m_q - i \gamma^\perp \partial_\perp + g_s \gamma^\perp A_\perp\right) \Psi^+,
\end{equation}
where \(m_q\) is the quark mass, $g_s$ is the coupling constant, and \(A_\mu\) is the gluon field. Herein, summation over all repeated indices is implied by default.

For the DVCS process (\(ep \rightarrow e'p'\gamma\)) associated with GPDs, we adopt a symmetric frame for the initial and final proton momenta. In this frame, the four-momenta of the incoming and outgoing protons are given by
\begin{align}
    P &= \bigg( (1+\xi)\bar{P}^+,-\frac{\vec{\Delta}^{\perp}}{2},\frac{M^2 + (\vec{\Delta}^\perp)^2/4}{(1+\xi)\bar{P}^+} \bigg) , \\
    P' &= \bigg( (1-\xi)\bar{P}^+,\frac{\vec{\Delta}^{\perp}}{2},\frac{M^2 + (\vec{\Delta}^\perp)^2/4}{(1-\xi)\bar{P}^+} \bigg) , \\
    \Delta &= P' - P = \bigg( -2\xi \bar{P}^+ , \vec{\Delta}^\perp , \frac{t - (\vec{\Delta}^\perp)^2}{2\xi \bar{P}^+} \bigg) ,
\end{align}
where \(\bar{P}=(P+P')/2\) is the average momentum of the proton, \(\Delta\) is the momentum transfer, \(\xi = -\Delta^{+}/(2\bar{P}^{+})\) is the skewness, \(M\) is the proton mass, and \(t\equiv -\Delta^2\) is the squared momentum transfer.

Based on the definitions of twist and Dirac matrices, the physical observables for the twist-3 case correspond to the interference terms \(\bar{\Psi}^+ \Gamma \Psi^-\) and \(\bar{\Psi}^- \Gamma \Psi^+\). By contrast, the twist-2 case is characterized by \(\bar{\Psi}^+ \Gamma \Psi^+\), and the twist-4 case by \(\bar{\Psi}^- \Gamma \Psi^-\). 
Using this general expression, the GPDs can be computed through the parametrization of Ref.~\cite{meissner2009generalized}. Several parametrization schemes for GPDs have been developed~\cite{PhysRevD.98.014038,guo2021novel,belitsky2001twist,belitsky2002theory,PhysRevD.88.014041,PhysRevD.100.096021}, among which the scheme of Ref.~\cite{kiptily2004genuine} is the most extensively studied and is directly related to the present work~\cite{PhysRevD.98.014038} (see Appendix. \ref{appb}). The full overlap expressions for all twist-3 GPDs are provided in Appendix~\ref{appa}. For the corresponding overlap formulas truncated to the lowest Fock sector, we refer the reader to our  earlier study~\cite{zhang2024twist}.

\section{\label{sec:level1}Numerical Results and Discussion} \label{sec4}
In this section, we present numerical results for zero-skewness twist-3 GPDs calculated via the BLFQ method. The key distinction between this work and previous studies~\cite{zhang2024twist} lies in the inclusion of two Fock sectors, \(|qqq\rangle + |qqqg\rangle\), which incorporates gluon contributions—specifically, the \(q\)-\(g\)-\(q\) interaction. We further analyze the properties of these twist-3 GPDs, including sum rules, spin-orbit correlations, and OAM. Note that properties such as time-reversal symmetry have been discussed in prior work and are not revisited herein.

The inclusion of two Fock sectors substantially increases the complexity of the overlap representation, primarily due to the emergence of the \(q\)-\(g\)-\(q\) correlator. Each twist-3 GPD consists of three distinct components: the \(qqq\) component, which arises from the overlap between the states \(\langle qqq|\) and \(|qqq \rangle\); the \(qqqg\) component, which originates from the overlap between \(\langle qqqg|\) and \(|qqqg\rangle\); and the ``genuine" (abbreviated as ``\(gen\)") component, which accounts for cross-overlap contributions, i.e., between \(\langle qqq|\) and \(|qqqg \rangle\), and between \(\langle qqqg|\) and \(|qqq \rangle\).

\begin{figure*}[htbp]
    \centering
    \subfloat[ ]{\includegraphics[scale=0.65]{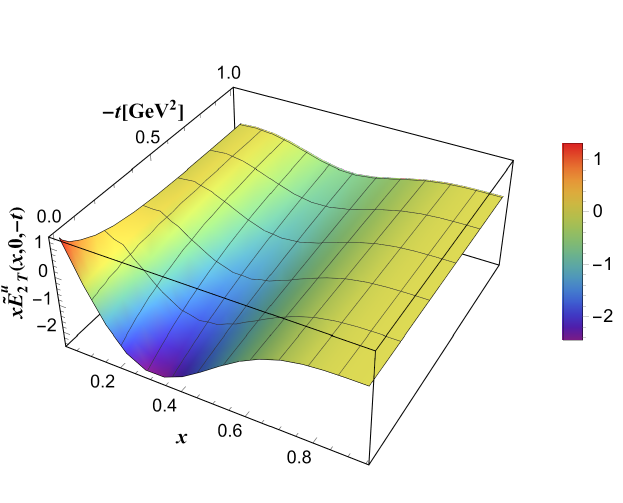}} \hspace{.3in}
    \subfloat[ ]{\includegraphics[scale=0.65]{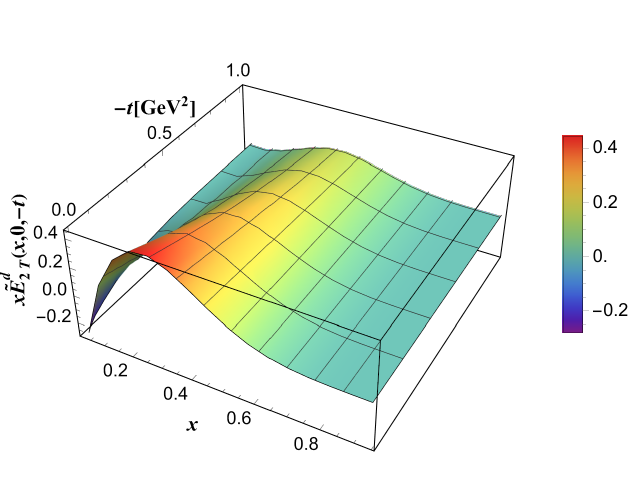}} \vspace{-.3in} \\
    \subfloat[ ]{\includegraphics[scale=0.65]{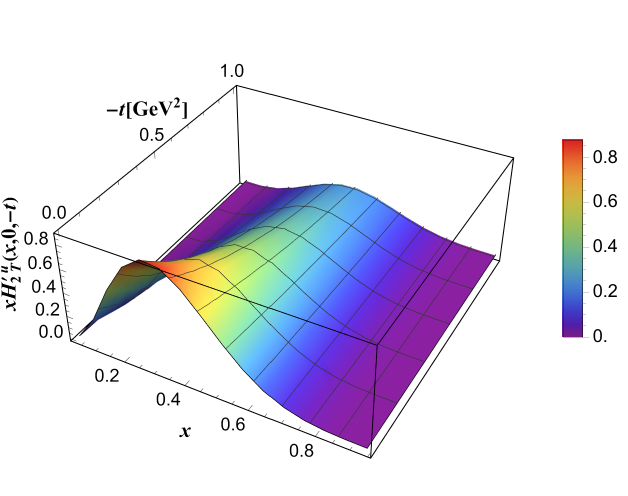}} \hspace{.3in}
    \subfloat[ ]{\includegraphics[scale=0.65]{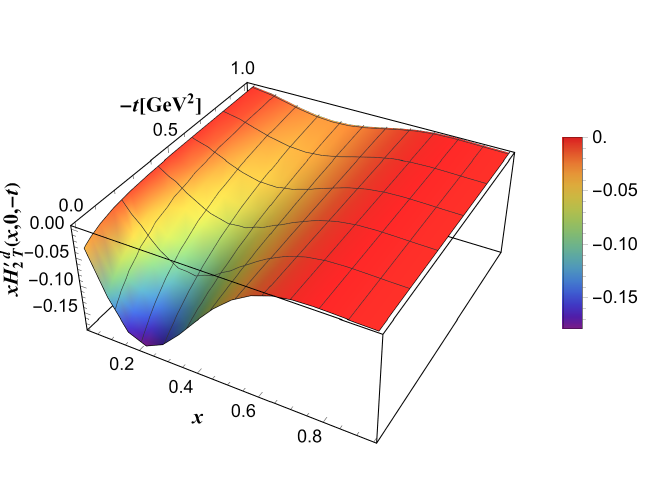}} \vspace{-.3in} \\
    \subfloat[ ]{\includegraphics[scale=0.65]{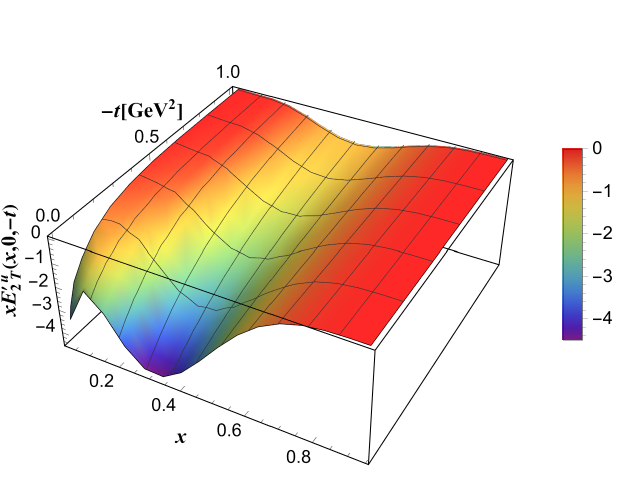}} \hspace{.3in}
    \subfloat[ ]{\includegraphics[scale=0.65]{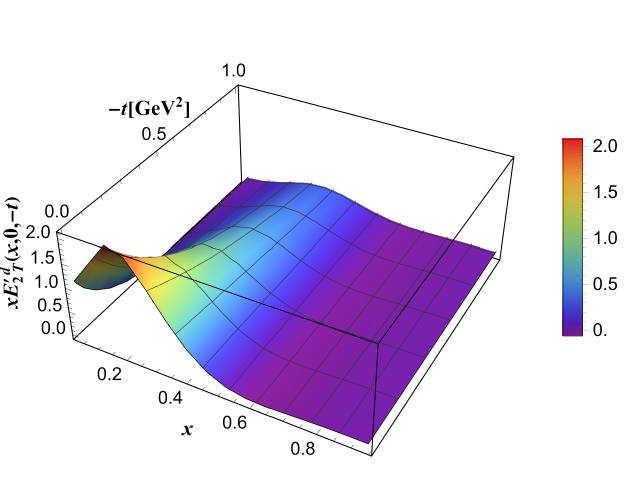}} \vspace{-.3in} \\
    \subfloat[ ]{\includegraphics[scale=0.65]{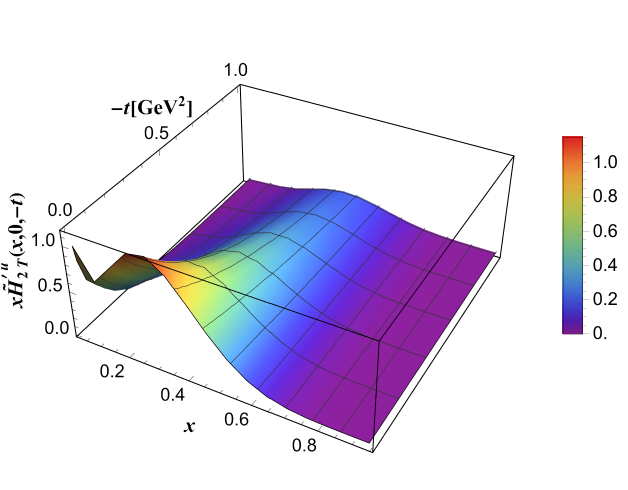}} \hspace{.3in}
    \subfloat[ ]{\includegraphics[scale=0.65]{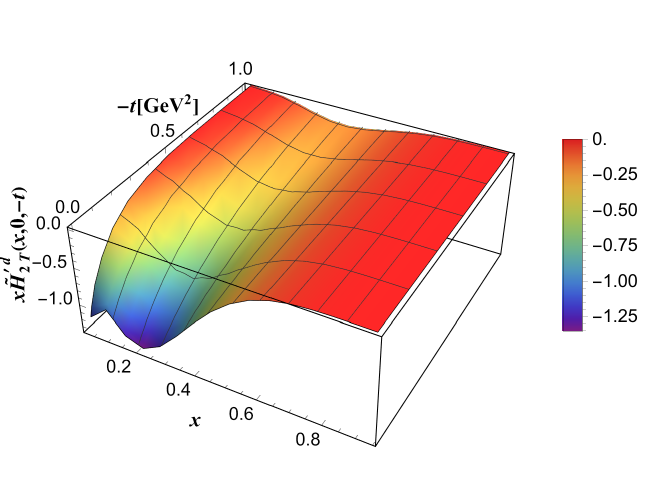}}
    \captionsetup{justification=raggedright}
    \caption{3D Plots of the twist-3 quark GPDs $x\tilde{E}_{2T}, xH_{2T}^\prime, xE_{2T}^\prime,$ and $x\tilde{H}_{2T}^\prime$ in the proton, evaluated with $N_{\mathrm{max}}=9$ and $K=16.5$. The left column panels \{(a), (c), (e), (g)\} correspond to the $u$ quark and the right column panels \{(b), (d), (f), (h)\} correspond to the $d$ quark on the flavor level. The flavor level distributions are given by $X_{flavor}^u =2 X_{quark}^u$ and $X_{flavor}^d=X_{quark}^d$, where $X$ stands for all the GPDs.}
    \label{plot3d}
\end{figure*}

\begin{figure*}[htbp]
    \centering
    \subfloat[ ]{\includegraphics[scale=0.85]{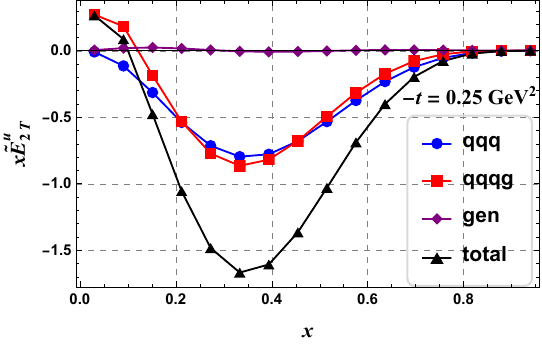}} \hspace{.3in}
    \subfloat[ ]{\includegraphics[scale=0.85]{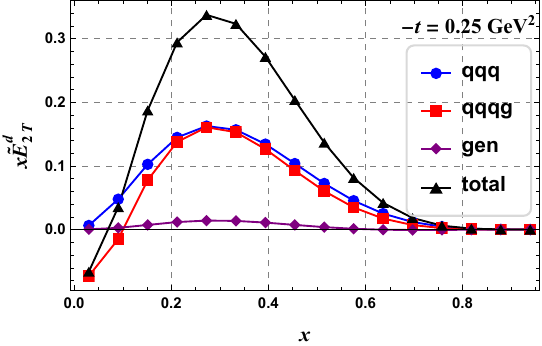}} \\
    \subfloat[ ]{\includegraphics[scale=0.85]{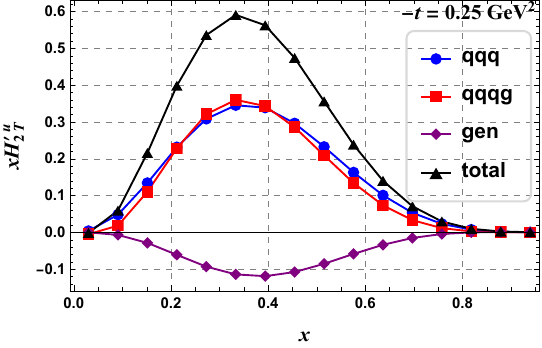}} \hspace{.3in}
    \subfloat[ ]{\includegraphics[scale=0.85]{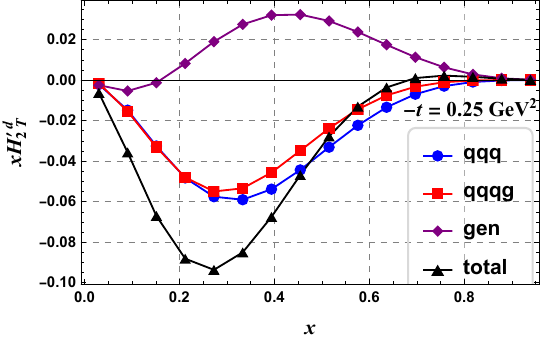}} \\
    \subfloat[ ]{\includegraphics[scale=0.85]{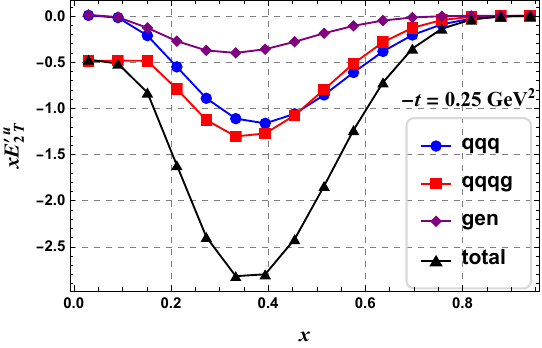}} \hspace{.3in}
    \subfloat[ ]{\includegraphics[scale=0.85]{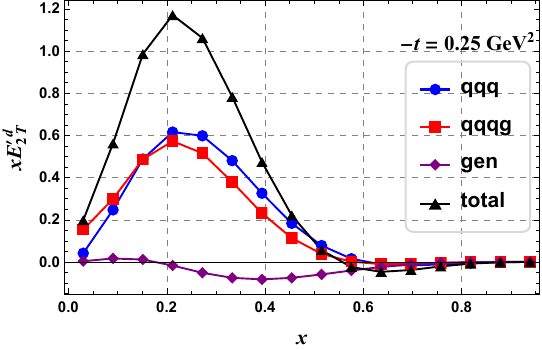}} \\
    \subfloat[ ]{\includegraphics[scale=0.85]{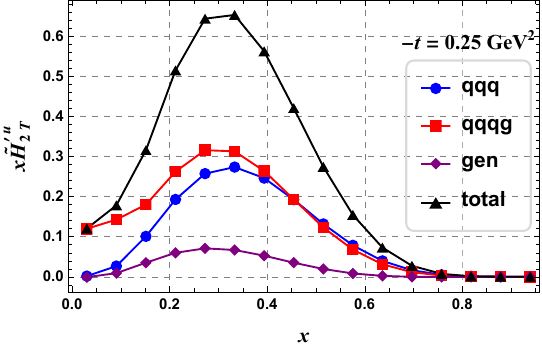}} \hspace{.3in}
    \subfloat[ ]{\includegraphics[scale=0.85]{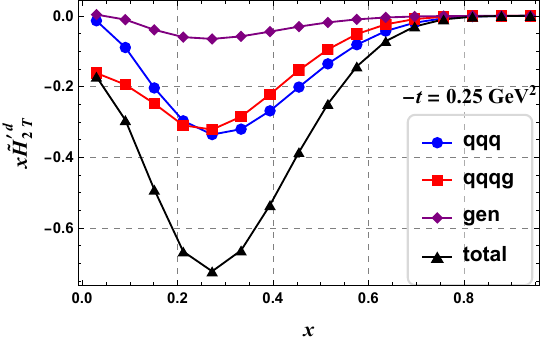}}
    \captionsetup{justification=raggedright}
    \caption{2D Plots of the twist-3 quark GPDs $x\tilde{E}_{2T}, xH_{2T}^\prime, xE_{2T}^\prime,$ and $x\tilde{H}_{2T}^\prime$ in the proton at fixed $-t=0.25$ GeV$^2$, evaluated with $N_{\mathrm{max}}=9$ and $K=16.5$. The left column panels \{(a), (c), (e), (g)\} correspond to the $u$ quark and the right column panels \{(b), (d), (f), (h)\} correspond to the $d$ quark on the flavor level. The flavor level distributions are given by $X_{flavor}^u =2 X_{quark}^u$ and $X_{flavor}^d=X_{quark}^d$, where $X$ stands for all the GPDs.}
    \label{plotfixt}
\end{figure*}

\begin{figure*}[htbp]
    \centering
    \subfloat[ ]{\includegraphics[scale=0.85]{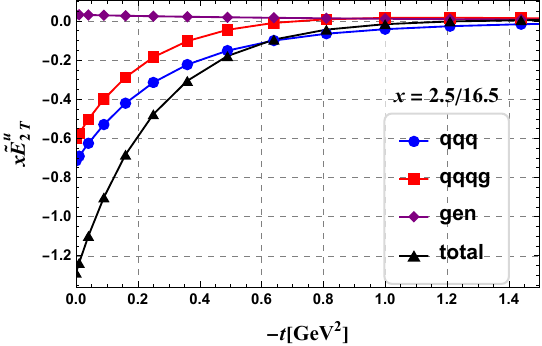}} \hspace{.3in}
    \subfloat[ ]{\includegraphics[scale=0.85]{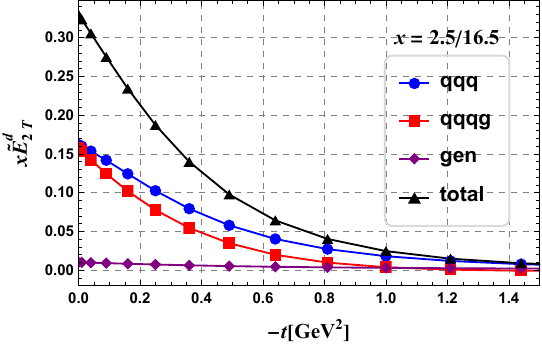}} \\
    \subfloat[ ]{\includegraphics[scale=0.85]{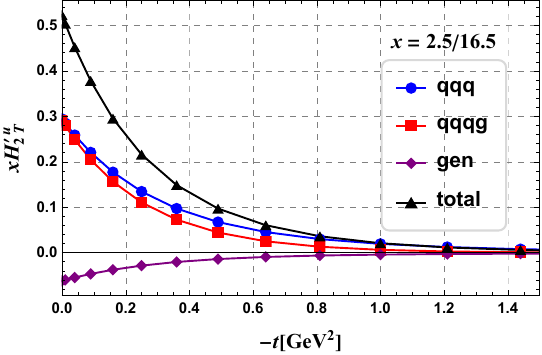}} \hspace{.3in}
    \subfloat[ ]{\includegraphics[scale=0.85]{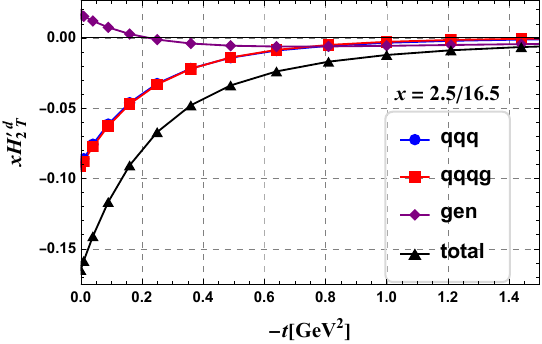}} \\
    \subfloat[ ]{\includegraphics[scale=0.85]{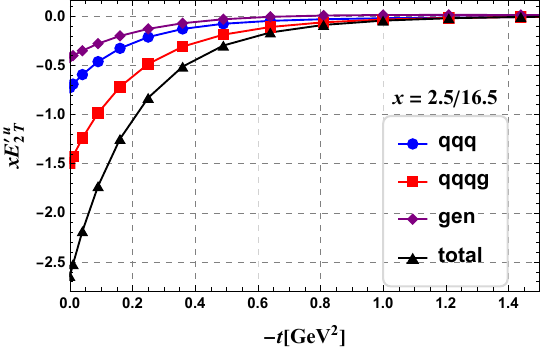}} \hspace{.3in}
    \subfloat[ ]{\includegraphics[scale=0.85]{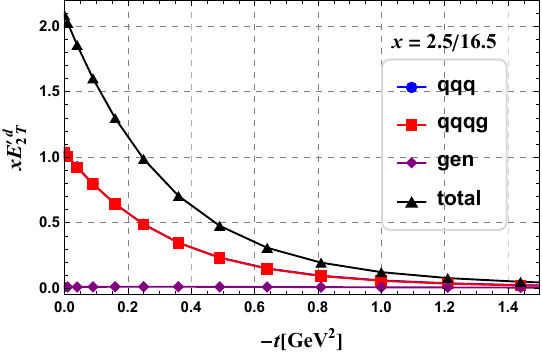}} \\
    \subfloat[ ]{\includegraphics[scale=0.85]{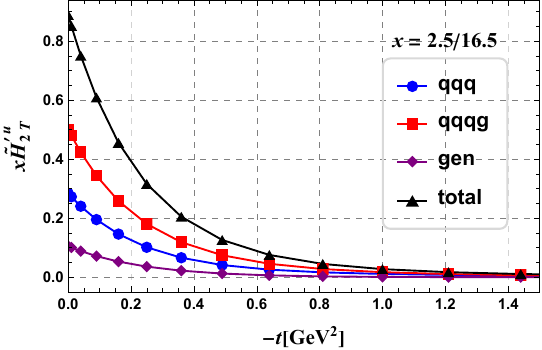}} \hspace{.3in}
    \subfloat[ ]{\includegraphics[scale=0.85]{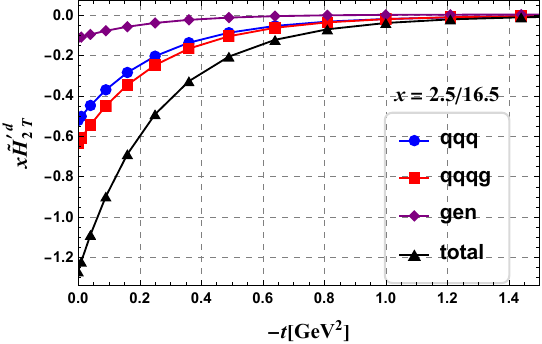}}
    \captionsetup{justification=raggedright}
    \caption{2D Plots of the twist-3 quark GPDs $x\tilde{E}_{2T}, xH_{2T}^\prime, xE_{2T}^\prime,$ and $x\tilde{H}_{2T}^\prime$ in the proton at fixed $x=2.5/16.5$, evaluated with $N_{\mathrm{max}}=9$ and $K=16.5$. The left column panels \{(a), (c), (e), (g)\} correspond to the $u$ quark and the right column panels \{(b), (d), (f), (h)\} correspond to the $d$ quark on the flavor level. The flavor level distributions are given by $X_{flavor}^u =2 X_{quark}^u$ and $X_{flavor}^d=X_{quark}^d$, where $X$ stands for all the GPDs.}
    \label{plotfixx}
\end{figure*}

\subsection{\label{sec:level2}Twist-3 GPDs}
For all results presented here, the rotational symmetry between \(\gamma_1\) and \(\gamma_2\) is preserved—not only for each individual component but also for their sum. For simplicity, \(\gamma_1\) is used for calculations. Given that \(\xi=0\) is fixed in this work, the GPDs are presented as functions of \(x\) and \(-t\). In the zero-skewness limit, eight of the twist-3 GPDs—\(H_{2T}\), \(E_{2T}\), \(\tilde{H}_{2T}\), \(\tilde{E}'_{2T}\), \(\tilde{H}_2\), \(H'_2\), \(E'_2\), and \(\tilde{E}'_2\)—are found to be consistent with zero within our numerical uncertainty.

Using the overlap representation and the LFWFs obtained by solving the BLFQ framework, we calculate all the twist-3 GPDs. For brevity, we present only the four most extensively studied chiral-even GPDs: these are parameterized from the vector current (\(\gamma^\perp\)) and pseudo-vector current (\(\gamma^\perp \gamma_5\)), namely \(\tilde{E}_{2T}\) (from vector current) and \(H_{2T}^\prime, E_{2T}^\prime, \tilde{H}_{2T}^\prime\) (from pseudo-vector current). These GPDs have been shown to relate to \(G_i\) and \(G_i^\prime\) (where \(i=1,4\))~\cite{PhysRevD.98.014038}. Each two-dimensional subplot distinguishes contributions from the \(qqq\) sector (blue solid circles), \(qqqg\) sector (red solid squares), genuine (\(gen\)) contribution (purple solid diamonds), and total GPD (black solid triangles); since the magnitudes of these curves are excessively large, we multiply by \(x\) to suppress their values.


In Fig.~\ref{plot3d}, we present the three-dimensional structure of the twist-3 GPDs in the zero-skewness limit. The results are plotted up to \(-t=1.0\) GeV$^2$, which focuses on the region with meaningful structural features. Beyond the upper limit of the depicted range, the magnitude maintains the depicted trends. That is, for \(-t > 1.0\) GeV$^2$ (not shown in the figures), the relevant quantities exhibit a featureless falloff to zero.

For \(x\tilde{E}_{2T}\), the region of small \(x\) and small \(-t\) shows a modest peak with a sign opposite to that in the rest of the kinematic domain; moreover, the \(u\) and \(d\) quark contributions carry opposite signs throughout the full region. In the case of \(xH^\prime_{2T}\), the distribution is smooth across all kinematics. This behavior arises because its overlap representation does not contain \(\Delta\) in the denominator, a feature that also makes it the only GPD that remains finite in the \(\Delta \rightarrow 0\) limit (corresponding to the twist-3 parton distribution function \(g_T(x)\)). 

For \(xE^\prime_{2T}\) and \(x\tilde{H}^\prime_{2T}\), the two distributions exhibit similar shapes but opposite signs—both between \(u\) and \(d\) quarks and between the two GPDs for a given quark flavor. Both GPDs also display fluctuations in the small \(x\) and small \(-t\) region, with signs consistent with their behavior in the rest of the domain. The peak observed at small \(x\) and small \(-t\) for all of the above GPDs originate from the inclusion of the nonvalence Fock sector with a dynamical gluon. Furthermore, owing to discretization and truncation effects in the longitudinal direction, we are not able to obtain reliable distributions at even smaller values of \(x\).

In Fig. \ref{plotfixt}, we fix \(-t=0.25\) GeV\(^2\) and plot the four GPDs as functions of \(x\). All GPDs show the same trend, with one prominent peak. The \(gen\) contribution is much smaller than the other components but remains non-negligible. Most GPDs fluctuate in the small \(x\) region: the \(qqqg\) sector contribution is particularly notable here, dominating in the small \(x\) region and nearly equal to the total GPD (except for \(H_{2T}^\prime\)). This behavior can be attributed to their overlap expressions, where division by \(\Delta\) amplifies contributions at small \(-t\). For the remaining \(x\) range, the \(qqq\) and \(qqqg\) sector contributions are nearly equal, while the \(gen\) contribution is minimal. The peaks of all GPDs cluster in the range \(0.2<x<0.4\), and all tend to zero in the large \(x\) regime.

In Fig. \ref{plotfixx}, we fix \(x=2.5/16.5\) and plot the four GPDs as functions of \(-t\). In subplots (a) and (b), the \(qqq\) sector contribution is slightly larger than that of the \(qqqg\) sector, while the \(gen\) contribution remains small. In subplots (c) and (d), the \(qqq\) and \(qqqg\) sector contributions are comparable. In subplot (e), the \(qqqg\) sector contribution is nearly twice that of the \(qqq\) sector, while in subplot (f), the two are nearly equal. In subplots (g) and (h), the \(qqqg\) sector contribution is larger than that of the \(qqq\) sector. All curves show a decreasing trend as \(-t\) increases.

\begin{figure*}[htbp]
    \centering
    \subfloat{\includegraphics[scale=0.85]{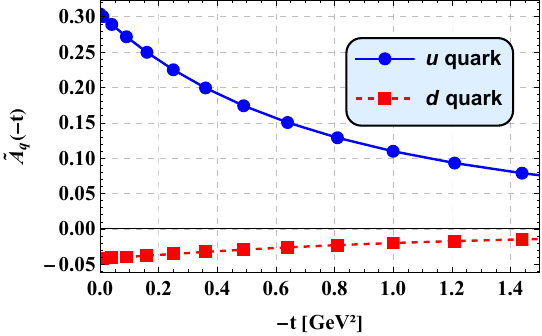}} \hspace{.3in}
    \subfloat{\includegraphics[scale=0.85]{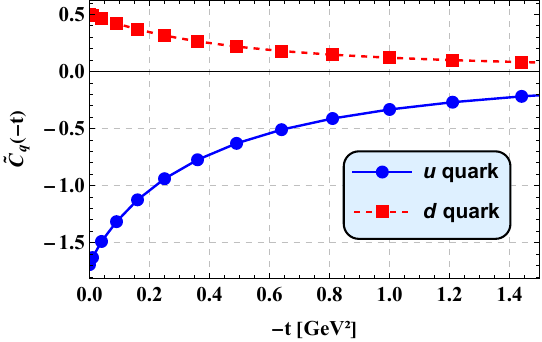}} \\
    \subfloat{\includegraphics[scale=0.85]{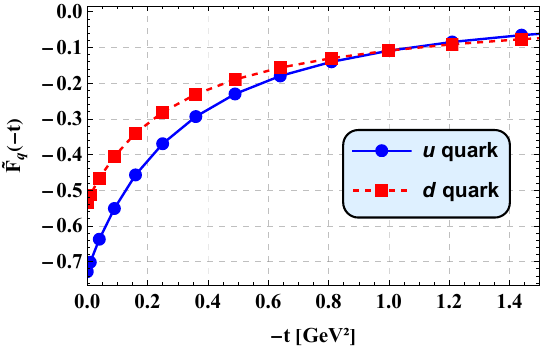}}
    \captionsetup{justification=raggedright}
    \caption{Plots of the twist-3 FFs $\tilde{A}_q, \tilde{C}_q$ and $\tilde{F}_q$ in the proton, evaluated with $N_{\mathrm{max}}=9$ and $K=16.5$. The blue circle solid line corresponds to the $u$ quark and the red square dotted line corresponds to the $d$ quark on the quark level.} 
    \label{plotffs}
\end{figure*}

Note that the sign reversal between the $u$ and $d$ distributions naturally reflects the spin-flavor structure of the nucleon in our model, analogous to the expectations from SU(6) symmetry and our previous light-front studies of quark helicities~\cite{Xu:2021wwj,Xu:2023nqv,zhang2024twist}. Twist-3 GPDs, however, lack a simple parton interpretation, and the origin of their sign reversal is not yet fully understood. It can nevertheless be interpreted as arising from the interplay between quark OAM and spin correlations.

\subsection{\label{sec:level2}Spin--orbit correlation}
The parity-odd partner of the quark energy-momentum tensor operator, \(\hat{T}^{\mu \nu}_{q5}\), is closely related to quark spin-orbit correlations and has been investigated in Ref.~\cite{lorce2014spin}. This operator is defined as
\begin{equation}
\hat{T}^{\mu \nu}_{q5} = \frac{1}{2} \bar{\psi}\gamma^\mu\gamma_5 i\overleftrightarrow{D^\nu}\psi,
\end{equation}
where \(\overleftrightarrow{D^\nu} = \overrightarrow{D^\nu} - \overleftarrow{D^\nu}\) denotes the symmetric covariant derivative (with \(D^\nu = \partial^\nu - igA^\nu\) as the covariant derivative). To characterize the matrix elements of \(\hat{T}^{\mu \nu}_{q5}\) in nucleon states, it is parametrized in terms of five FFs
\begin{equation}
    \langle p',s'| \hat{T}^{\mu \nu}_{q5} |p,s \rangle = \bar{u}(p',s') \Gamma^{\mu \nu}_{q 5} u(p,s),
\end{equation}
with
\begin{align}
    \Gamma^{\mu \nu}_{q 5} &= \frac{\bar{P}^{\{\mu}\gamma^{\nu\}}\gamma_5}{2}\tilde{A}_q(-t) + \frac{\bar{P}^{\{\mu}\Delta^{\nu\}}\gamma_5}{4M}\tilde{B}_q(-t) \notag \\
    &+ \frac{\bar{P}^{[\mu}\gamma^{\nu]}\gamma_5}{2}\tilde{C}_q(-t) + \frac{\bar{P}^{[\mu}\Delta^{\nu]}\gamma_5}{4M}\tilde{D}_q(-t) \notag \\
    &+ Mi\sigma^{\mu\nu}\gamma_5\tilde{F}_q(-t).
\end{align}

The FFs $\tilde{A}_q(-t)$ and $\tilde{B}_q(-t)$ are associated with the twist-2 contribution obtained by selecting the $\mu\nu = ++$ component of the energy$-$momentum tensor. In this case, one finds:
\begin{align}
    \int dxx\tilde{H}_q(x,\xi,-t) &= \tilde{A}_q(-t), \\
    \int dxx\tilde{E}_q(x,\xi,-t) &= \tilde{B}_q(-t). 
\end{align}

Taking the twist-3 component $\mu\nu=j+$ with $j=1,2$, the following relations are obtained~\cite{lorce2014spin}
\begin{align}
    \int dx x \tilde{G}_1^q (x,\xi,-t) =& -\frac{1}{2} \bigg( \tilde{B}_q(-t) + \tilde{D}_q(-t) \bigg), \\
    \int dx x \tilde{G}_2^q (x,\xi,-t) =& -\frac{1}{2} \bigg( \tilde{A}_q(-t) + \tilde{C}_q(-t) \bigg) \notag \\ 
    &+ (1-\xi^2)\tilde{F}_q(-t), \\
    \int dx x \tilde{G}_3^q (x,\xi,-t) =& -\frac{\xi}{2} \tilde{F}_q(-t), \\
    \int dx x \tilde{G}_4^q (x,\xi,-t) =& -\frac{1}{2} \tilde{F}_q(-t),
\end{align}
where  \(\tilde{G}_i\) are related to \(H_{2T}^\prime\), \(E_{2T}^\prime\), \(\tilde{H}_{2T}^\prime\), and \(\tilde{E}_{2T}^\prime\)~\cite{PhysRevD.98.014038,zhang2024twist} (see Appendix. \ref{appb}).

We present the FFs \(\tilde{A}_q(-t)\), \(\tilde{C}_q(-t)\), and \(\tilde{F}_q(-t)\) in Fig.~\ref{plotffs}. We observe that $\tilde{A}_q(-t)$ and $\tilde{C}_q(-t)$ exhibit similar shapes but have opposite signs---both between $u$- and $d$-quarks and between the two form factors for a given quark flavor. Moreover, the $u$-quark contribution is significantly larger than that of the $d$-quark. In contrast, $\tilde{F}_q(-t)$ is negative for both flavors and displays a similar overall behavior.

Since we focus on extracting the GPDs at $\xi = 0$, the distributions $\tilde{E}$ and $\tilde{G}_1$ lie beyond the scope of this work. Consequently, the corresponding form factors $\tilde{B}_q(-t)$ and $\tilde{D}_q(-t)$ are not determined in the present analysis. For clarity, these form factors could, in principle, be obtained by performing calculations at nonzero $\xi$, where their contributions can be isolated through the full GPD parametrization.

Several well-established relations involving twist-3 GPDs in the forward limit are worth highlighting. The quark OAM in the nucleon is defined, in the limit \(\Delta \rightarrow 0\), as \cite{kiptily2004genuine}
\begin{equation}
    L_z^q = - \lim_{\Delta \rightarrow 0} \int dx\, x\, G_2^q(x,0,\Delta^2_\perp),
\end{equation}
where the explicit expression for $ G_2^q$ is provided in Appendix~\ref{appb}. This physical quantity can also be calculated directly from twist-2 GPDs, as defined in Refs.~\cite{ji1997gauge,hatta2012twist,JAFFE1990509}. Within our BLFQ approach at the model scale, we obtain $L^u_z = 0.107$ ($L^d_z = 0.0752$) from twist-3 distributions and $L^u_z = 0.0817$ ($L^d_z = 0.0264$) from twist-2 GPDs. This difference may arise from the violation of the Burkhardt-Cottingham sum rule \cite{burkhardt1970sum} ($\int dx g_1(x)=\int dx g_T(x)$) in our results, which originates from the explicit breaking of rotational invariance in our approach. Owing to Fock sector truncation, the full QCD Hamiltonian interactions cannot be incorporated, and separate confining potentials are thus introduced for the longitudinal and transverse directions, whereas the BC sum rule itself is viewed as a consequence of rotational symmetry~\cite{Jaffe:1997vlv}. A similar deviation has also been reported in Ref. \cite{tan2024chiral}.

The total quark spin contribution $J_z^q$ can also be related to twist-3 GPDs via~\cite{guo2021novel}
\begin{equation}
    J^q_z = \int dx \left( x G_{q,3}(x,0,0) - \frac{1}{2} g_1^q(x) \right),
\end{equation}
where $g_1^q$ is the twist-2 helicity PDF, and $G_{q,3}$ (defined in Ref.~\cite{guo2021novel}) can be converted to $G_{q,3} = -\tilde{E}^q_{2T}$. Using this relation, at the model scale, we find $J_z^u = 0.443$ ($J_z^d = -0.0801$) from twist-3 distributions and $J_z^u = 0.468$ ($J_z^d = -0.0312$) from twist-2 distributions.

For the quark spin-orbit correlation, it is expressed as~\cite{lorce2014spin}
\begin{equation}
    C_z^q = -\int dx x \big( \tilde{G}_2^q(x,0,0) + 2\tilde{G}_4^q(x,0,0) \big).
\end{equation}
We list our results in Table \ref{cztabel} for comparison with other theoretical approaches, including the naive quark model (NQM), the light-front constituent quark model (LFCQM), and the light-front chiral quark-soliton model (LF$\chi$QSM) \cite{lorce2011quark}, the Leader-Sidorov-Stamenov (LSS) analysis \cite{leader2010determination}, and the spectator diquark model \cite{tan2024chiral}. The negative signs of the results indicate that the quark spin and Ji OAM are anti-correlated on average \cite{lorce2014spin}.


\begin{table}[h!]
\caption{\label{cztabel}
Comparison of $C_z^u$ and $C_z^d$ values from different theoretical approaches.}
\begin{ruledtabular}
\begin{tabular}{lcc}
Theoretical Approach & $C_z^u$ & $C_z^d$ \\
\colrule
BLFQ (model scale, this work) & $-0.761$ & $-0.515$ \\
NQM, LFCQM, LF$\chi$QSM \cite{lorce2011quark} & $\approx -0.8$ & $\approx -0.55$ \\
LSS ($\mu^2=1$ GeV$^2$) \cite{leader2010determination} & $\approx -0.9$ & $\approx -0.53$ \\
Spectator Diquark Model \cite{tan2024chiral} & $-0.775$ & $-0.586$ \\
\end{tabular}
\end{ruledtabular}
\end{table}


\section{\label{sec:level1}Conclusion} \label{sec5}
In this work, we have computed all twist-3 GPDs of the proton in the zero-skewness limit within the BLFQ framework, presenting the four non-vanishing chiral-even distributions for both \(u\) and \(d\) quarks. Building upon our previous study, we have extended the Fock-space truncation to include a dynamical gluon, thereby incorporating contributions from the \(q\text{--}g\text{--}q\) correlator---i.e., the genuine twist-3 component. The light-front wave functions used in the overlap representation are obtained by diagonalizing the proton 
light-front Hamiltonian constructed for the \(|qqq\rangle\) and \(|qqqg\rangle\) Fock sectors, supplemented with a three-dimensional confinement potential.

Our calculations indicate that the genuine twist-3 contribution, while relatively small, is nevertheless non-negligible; in certain cases, it even induces a sign opposite to that of the non-genuine component. Qualitatively, our results show good overall consistency with those of Ref.~\cite{jain2024deciphering}, except for the \(x\tilde{E}_{2T}^d\) distribution and in the region of small \(x\) and small \(-t\). In addition, we have examined several key twist-3--related quantities—including the quark OAM, the total quark contribution to the nucleon spin, and the quark spin–orbit correlation—and many of these are in reasonable agreement with predictions from other model studies~\cite{lorce2014spin,lorce2013wilson,leader2010determination,tan2024chiral}.

In future work, we plan to extend our analysis to twist-3 GPDs at nonzero skewness and, more broadly, to the twist-3 distributions of gluons and sea quarks. The nonzero-skewness sector is particularly important, as these twist-3 GPDs enter the twist-3 DVCS amplitudes~\cite{guo2022twist} and will be probed in forthcoming measurements at the EIC and EicC~\cite{chavez2022accessing}. With continued development of the BLFQ framework, we expect to improve the precision of our calculations, address remaining discrepancies in sum-rule relations, and provide more reliable predictions relevant to understanding the proton spin puzzle.

\begin{acknowledgments}
We thank Jiangshan Lan, Zhi Hu, Zhimin Zhu and Jiatong Wu for many helpful discussions and useful advises. C. M. is supported by new faculty start up funding the Institute of Modern Physics, Chinese Academy of Sciences, Grants No. E129952YR0. X. Z. is supported by the National Natural Science Foundation of China under Grant No.12375143, by new faculty startup funding by the Institute of Modern Physics, Chinese Academy of Sciences, by Key Research Program of Frontier Sciences, Chinese Academy of Sciences, Grant No. ZDBS-LY-7020, by the Natural Science Foundation of Gansu Province, China, Grant No. 20JR10RA067, by the Foundation for Key Talents of Gansu Province, by the Central Funds Guiding the Local Science and Technology Development of Gansu Province, Grant No. 22ZY1QA006, by international partnership program of the Chinese Academy of Sciences, Grant No. 016GJHZ2022103FN, by the Strategic Priority Research Program of the Chinese Academy of Sciences, Grant No. XDB34000000, and by National Key R$\&$D Program of China, Grant No. 2023YFA1606903. J. P. V. is supported by the Department of Energy under Grant No. DE-SC0023692. A portion of the computational resources were also provided by Dongjiang Yuan Intelligent Computing Center.
\end{acknowledgments}

\appendix

\section{Overlap representations for twist-3 GPDs} \label{appa}
We will present the overlap representations of all zero-skewness twist-3 GPDs in the following. For convenience, the following notations are adopted,
\begin{align}
    [dx]_n &= \prod_{i=1}^n \frac{d x_{i} d^2 \vec{k}_{i}}{(16 \pi^3)^n} 16\pi^3 \delta \left(1-\sum x_i \right) \notag \\ 
    &\times \delta^2 \left(\sum \vec{k}_i \right) \delta (x-x_1),
\end{align}
where the subscript 1 denotes the struck quark, and $[\Gamma]=\bar{u} (p',\lambda') \Gamma u (p,\lambda)$ encapsulates the helicity combinations of the struck quark. Additionaly, $\psi^{\Lambda}_n$ signifies the LFWF $\psi^{\Lambda}_{i=1,\cdots,n}(x_i,p_i,\lambda_i)$, with $\Lambda$ representing the proton helicity and $\lambda_i$ represents the quark helicity. The constraint for spectators $\delta^{\lambda'_2}_{\lambda_2} \delta^{\lambda'_3}_{\lambda_3} \delta^{\lambda'_4}_{\lambda_4}$ is implicitly assumed. In the overlap expressions, the symbol $\Delta$ refers to the two-dimensional complex representation, i.e., $\Delta = \Delta_{1} + i\Delta_{2}$.

Each overlap representation is divided into three parts. For the $|qqq \rangle$ sector, we have
\begin{align}
    \mathscr{F}^{qqq}_{\Gamma, \Lambda^\prime \Lambda} =& \int [dx]_3 \sum_{\lambda_1^\prime, \lambda_1} \left( \frac{p_{1 \perp}}{x_1} [\Gamma \gamma^+ \gamma^\perp] + \frac{p_{1 \perp}^\prime}{x_1} [\gamma^\perp \gamma^+ \Gamma] \right. \notag \\
    & \left. + \frac{m_q}{x_1} [\Gamma \gamma^+] + \frac{m_q}{x_1} [\gamma^+ \Gamma] \right) \psi^{\Lambda^\prime \star}_{3} \psi^{\Lambda}_{3}.
\end{align}
A similar expression applies to the $|qqqg \rangle$ sector,
\begin{align}
    \mathscr{F}^{qqqg}_{\Gamma, \Lambda^\prime \Lambda} =& \int [dx]_4 \sum_{\lambda_1^\prime, \lambda_1} \left( \frac{p_{1 \perp}}{x_1} [\Gamma \gamma^+ \gamma^\perp] + \frac{p_{1 \perp}^\prime}{x_1} [\gamma^\perp \gamma^+ \Gamma] \right. \notag \\
    & \left. + \frac{m_q}{x_1} [\Gamma \gamma^+] + \frac{m_q}{x_1} [\gamma^+ \Gamma] \right) \psi^{\Lambda^\prime \star}_{4} \psi^{\Lambda}_{4}.
\end{align}
Regarding the interference term $gen$, which incorporates the $q$-$g$-$q$ interaction, we have
\begin{align}
    \mathscr{F}^{gen}_{\Gamma, \Lambda^\prime \Lambda} =& \int [dx]_4 \sum_{\lambda_1^\prime, \lambda_1, \lambda_g} \frac{g_s}{\sqrt{x_g}} \left\{ \left( \frac{\epsilon^j}{x_1^\prime} [\Gamma \gamma^+ \gamma^j] \right. \right. \notag \\
    & + \left. \left. \frac{\epsilon^j}{x_1} [\gamma^j \gamma^+ \Gamma] \right) \right. \psi^{\Lambda^\prime \star}_{3} \psi^{\Lambda}_{4} + \left( \frac{\epsilon^{j \star}}{x_1} [\gamma^j \gamma^+ \Gamma] \right. \notag \\
    & + \left. \left. \frac{\epsilon^{j \star}}{x_1^\prime} [\Gamma \gamma^+ \gamma^j] \right) \psi^{\Lambda^\prime \star}_{4} \psi^{\Lambda}_{3}  \right\},
\end{align}
where $\epsilon^\mu$ is the gluon polarization vector.

The total contribution is the sum of these three components,
\begin{equation}
    \mathscr{F}^{total}_{\Gamma, \Lambda^\prime \Lambda} = \mathscr{F}^{qqq}_{\Gamma, \Lambda^\prime \Lambda} + \mathscr{F}^{qqqg}_{\Gamma, \Lambda^\prime \Lambda} + \mathscr{F}^{gen}_{\Gamma, \Lambda^\prime \Lambda}.
\end{equation}

Using the expressions derived above, we can explicitly formulate the overlap representations for all twist-3 GPDs
\begin{align}
    H_{2} =& \frac{P^+}{M} \mathscr{F}^{total}_{1,\uparrow \uparrow}, \\
    E_{2} =& \frac{2 P^+}{\Delta^\star} \mathscr{F}^{total}_{1,\uparrow \downarrow}, \\
    \tilde{H}_{2} =& \frac{P^+}{M} \mathscr{F}^{total}_{\gamma_5,\uparrow \uparrow}, \\
    \tilde{E}_{2} =& -\frac{2 P^+}{\Delta^\star} \mathscr{F}^{total}_{\gamma_5,\uparrow \downarrow}, \\
    H_{2T}^j =& \frac{(-i)^{(j-1)} P^+}{Mi(\Delta + (-)^j \Delta^\star)} \left( \Delta \mathscr{F}^{total}_{\gamma^j,\uparrow \downarrow} + \Delta^\star \mathscr{F}^{total}_{\gamma^j, \downarrow \uparrow} \right), \\
    E_{2T}^j =& \frac{P^+}{\Delta_j} \left(\mathscr{F}^{total}_{\gamma^j,\uparrow \uparrow} + \mathscr{F}^{total}_{\gamma^j,\downarrow \downarrow}\right) \notag \\
    &+ \frac{2M P^+}{i^j \Delta_1 \Delta_2} \left(\mathscr{F}^{total}_{\gamma^j,\uparrow \downarrow} - (-)^j \mathscr{F}^{total}_{\gamma^j, \downarrow \uparrow}\right), \\
    \tilde{H}_{2T}^j =& -\frac{M P^+}{i^j \Delta_1 \Delta_2} \left( \mathscr{F}^{total}_{\gamma^j,\uparrow \downarrow} - (-)^j \mathscr{F}^{total}_{\gamma^j,\downarrow \uparrow} \right), \\
    \tilde{E}_{2T}^j =& \frac{2i^{(j-1)}P^+}{\Delta + (-)^j \Delta^\star} \left( \mathscr{F}^{total}_{\gamma^j,\uparrow \uparrow} - \mathscr{F}^{total}_{\gamma^j,\downarrow \downarrow} \right), \\
    H_{2T}^{\prime j} =& \frac{P^+}{2M\Delta_j} \left( \Delta \mathscr{F}^{total}_{\gamma^j \gamma_5,\uparrow \downarrow} + \Delta^\star \mathscr{F}^{total}_{\gamma^j \gamma_5, \downarrow \uparrow} \right), \\
    E_{2T}^{\prime j} =& \frac{2i^{(j-1)}P^+}{\Delta + (-)^j \Delta^\star} \left( \mathscr{F}^{total}_{\gamma^j \gamma_5,\uparrow \uparrow} + \mathscr{F}^{total}_{\gamma^j \gamma_5,\downarrow \downarrow} \right) \notag \\
    &+ \frac{2M P^+}{i^{(2-j)} \Delta_1 \Delta_2} \left((-)^{j} \mathscr{F}^{total}_{\gamma^j \gamma_5,\uparrow \downarrow} + \mathscr{F}^{total}_{\gamma^j \gamma_5, \downarrow \uparrow}\right), \\
    \tilde{H}_{2T}^{\prime j} =& \frac{(-)^{(j-1)} M P^+}{i^{(2-j)} \Delta_1 \Delta_2} \left( \mathscr{F}^{total}_{\gamma^j \gamma_5,\uparrow \downarrow} - (-)^j \mathscr{F}^{total}_{\gamma^j \gamma_5,\downarrow \uparrow} \right), \\
    \tilde{E}_{2T}^{\prime j} =& \frac{P^+}{\Delta_j} \left( \mathscr{F}^{total}_{\gamma^j \gamma_5,\uparrow \uparrow} - \mathscr{F}^{total}_{\gamma^j \gamma_5,\downarrow \downarrow} \right), \\
    H_{2}^\prime =& \frac{i P^+}{M} \mathscr{F}^{total}_{i\sigma^{12}\gamma_5,\uparrow \uparrow}, \\
    E_{2}^\prime =& \frac{2i P^+}{\Delta^\star} \mathscr{F}^{total}_{i\sigma^{12}\gamma_5,\uparrow \downarrow}, \\
    \tilde{H}_{2}^\prime =& \frac{P^+}{M} \mathscr{F}^{total}_{i\sigma^{+-}\gamma_5,\uparrow \uparrow}, \\
    \tilde{E}_{2}^\prime =& -\frac{2 P^+}{\Delta^\star} \mathscr{F}^{total}_{i\sigma^{+-}\gamma_5,\uparrow \downarrow},
\end{align}
where $M$ denotes the proton mass. All the numerical results presented in this work are calculated using these overlap representations, and the LFWF obtained from the BLFQ framework.

\section{Relations Between Different GPDs} \label{appb}
The relations between GPDs defined in Ref.~\cite{PhysRevD.98.014038} and in ~\cite{meissner2009generalized} are
\begin{align}
    H_{2T} =& 2\xi G_{4}, \\
    E_{2T} =& 2(G_{3} - \xi G_{4}), \\
    \tilde{H}_{2T} =& \frac{1}{2} G_{1}, \\
    \tilde{E}_{2T} =& -(H + E + G_{2}) + 2(\xi G_{3} - G_{4}), \\
    H'_{2T} =& \frac{-t}{4M^{2}} (\tilde{E} + \tilde{G}_{1}) + (\tilde{H} + \tilde{G}_{2}) - 2\xi \tilde{G}_{3}, \\
    E'_{2T} =& -(\tilde{E} + \tilde{G}_{1}) - (\tilde{H} + \tilde{G}_{2}) + 2(\xi \tilde{G}_{3} - \tilde{G}_{4}), \\
    \tilde{H}'_{2T} =& \frac{1}{2} (\tilde{E} + \tilde{G}_{1}), \\
    \tilde{E}'_{2T} =& 2(\tilde{G}_{3} - \xi \tilde{G}_{4}) .
\end{align}

The inversion of above equations is
\begin{align}
    G_{1} =& 2\tilde{H}_{2T}, \label{invst} \\
    G_{2} =& -(H + E) - \frac{1}{\xi} (1-\xi^{2}) H_{2T} + \xi E_{2T} - \tilde{E}_{2T}, \\
    G_{3} =& \frac{1}{2} (H_{2T} + E_{2T}), \\
    G_{4} =& \frac{1}{2\xi} H_{2T}, \\
    \tilde{G}_{1} =& -\tilde{E} + 2\tilde{H}'_{2T}, \\
    \tilde{G}_{2} =& -\tilde{H} + (1-\xi^{2}) H'_{2T} -\xi^{2} E'_{2T} - \frac{\Delta^{2}_{\perp}}{2M^{2}} \tilde{H}'_{2T} + \xi \tilde{E}'_{2T}, \\
    \tilde{G}_{3} =& -\frac{\xi}{2} (H'_{2T} + E'_{2T}) - \frac{\xi \bar{M}^{2}}{M^{2}} \tilde{H}'_{2T} + \frac{1}{2} \tilde{E}'_{2T}, \\
    \tilde{G}_{4} =& -\frac{1}{2} (H'_{2T} + E'_{2T}) - \frac{\bar{M}^{2}}{M^{2}} \tilde{H}'_{2T} \label{inved},
\end{align}
where $\bar{M}^{2} = M^{2} + t/4$.

The relations between GPDs defined in Ref.~\cite{guo2021novel} and this work are 
\begin{align}
    E_{2T} (x,\xi,-t) &= 2 G_{q,2} (x,\xi,-t), \label{jis1} \\
    H_{2T} (x,\xi,-t) &= G_{q,4} (x,\xi,-t), \label{jis2} \\
    \tilde{E}_{2T} (x,\xi,-t) &= 2\xi G_{q,2} (x,\xi,-t) - G_{q,3} (x,\xi,-t), \\
    \tilde{H}_{2T} (x,\xi,-t) &= G_{q,1} (x,\xi,-t). \label{jis4}
\end{align}

\bibliographystyle{unsrt}
\bibliography{main}

\end{document}